# Big Data analytics and Cognitive Computing – future opportunities for Astronomical research

M.A. Garrett[1,2]
[1]ASTRON, the Netherlands Institute for Radio Astronomy, Postbox 2, 7990 AA Dwingeloo, The Netherlands
[2]University of Leiden, Leiden Observatory, Postbus 9513, 2300 RA Leiden, The Netherlands.

E-mail: garrett@astron.nl

**Abstract.** The days of the lone astronomer with his optical telescope and photographic plates are long gone: Astronomy in 2025 will not only be multi-wavelength, but multi-messenger, and dominated by huge data sets and matching data rates. Catalogues listing detailed properties of billions of objects will in themselves require a new industrial-scale approach to scientific discovery, requiring the latest techniques of advanced data analytics and an early engagement with the first generation of cognitive computing systems. Astronomers have the opportunity to be early adopters of these new technologies and methodologies – the impact can be profound and highly beneficial to effecting rapid progress in the field. Areas such as SETI research might favourably benefit from cognitive intelligence that does not rely on human bias and preconceptions.

## 1. Introduction

Astronomical facilities, such as the new generation of radio telescopes e.g. LOFAR [1] are often considered as good examples of the type of Big Data generators that are rapidly emerging in a wide variety of different scientific fields. Telescope facilities like LOFAR can routinely produce data sets of ~ 100 TB per day, and new facilities such as the SKA [2] will generate object catalogues that contain billions of sources [3]. The limited ability to process these huge data sets and make sense of the complexity of the archived data products is already emerging as a major bottleneck in the scientific exploitation of these new instruments.

Initiatives such as DOME, the ASTRON IBM Center for Exascale Technology [4] are addressing some of these technical issues by investing significant effort in areas that include low-power (green) data streaming platforms, exascale storage solutions and the extensive implementation of nano-photonics. The challenge of the SKA forces much of the work to concentrate on realising systems that can deliver much higher performance, on a reduced footprint with lower power consumption. These efforts mostly focus on front-end hardware and back-end data challenges but there is a clear ambition to begin to extend this work into the post-processing domain via advanced Big Data analytics. In particular, the next generation of all-sky astronomical surveys are set to generate a total data archive that also approaches Petabyte scales. It is no longer possible to consider that a single astronomer, or a large scientific team, or even a global citizen scientist project can fully exploit the scientific potential of these data sets.

Astronomers are in a good position to benefit from the nature of their data and the extreme scale of their problem - telescopes like LOFAR and SKA make for interesting test beds for the development of novel techniques and advanced systems that can have additional value in many other fields, including those with significant societal benefit via commercial and industrial application. By helping to develop and become early adopters of these advances, astronomers have the opportunity to make great strides in fully exploiting advances in data analytics and cognitive computing in order to fully exploit the huge catalogues of multi-dimensional, multi-wavelength and multi-messenger data coming their way. The DOME collaboration is particularly well placed to further develop these ideas - in this paper we consider how these efforts might develop over the next 10-15 years.

## 2. Big Data and advanced analytics for astronomy

Big Data is usually defined in terms of the three V's – Volume, Velocity and Variety [5]. All three characteristics apply to astronomical data – the first two are obvious (large data sets and data streaming or data "in motion"), and the third "Variety" relates to the diversity of data type – in astronomy this can be the considered to relate to the myriad of different instruments, wavelength range/messenger type recorded, and data format. A fourth V, "Veracity" is important in all areas of analytics, and is well known to astronomers dealing with measurements that are inherently uncertain (due to limits of the instrumentation) or possibly corrupt (e.g. due to the presence of man-made interference in radio astronomy data sets).

While it is true that telescopes like LOFAR are prodigious generators of the 4 V's, the amount of data being generated around the planet is also quite staggering. It's estimated for example, that Google currently processes ~ 25 PB of data every day, and that ~ 3 EB of data are generated every day around the world. The Internet itself is an impressive beast with a total storage space of ~ 250 EB, total data rates of ~ 8TB/second and a total cost estimated at 5% of the world's total power requirements. Cloud storage systems already count for 1 EB of the total storage capacity, and this capacity is set to grow while the costs reduce. By the end of the decade, this data avalanche enhanced by the growth of sensor networks (of which LOFAR is one of the first multi-disciplinary examples) is expected to increase 30-fold.

Analysing the combination of traditional data sources with unstructured data (e.g. social media) is today's big business, and the investments being made by large commercial companies are significant. Many of the advances now being made via a variet of Big Data initiatives are likely to be applicable to astronomy – not only the distributed server hardware but also the underlying software infrastructure e.g. scalable open-source software such as Hadoop, supporting distributed processing of large data sets across clusters of commodity servers. In order to understand and predict the behavior of customers, forecast financial trends, or uncover operational insights that can increase efficiency, huge investments are being made in areas such as data analytics. Often, correlations and patterns in the data must be identified quickly – usually in real-time, if they are to realise their full commercial impact. Brute force correlations of all data types are leading to new and surprising insights that were not intuitively obvious from the start. Naturally, it's important not to confuse correlation with causality but a brute force approach, coupled with sophisticated pattern recognition algorithms can reveal hard-to-spot connections.

As the size of astronomical survey data sets is set to grow considerably over the next decade, an approach based on commercial data analytics may offer considerable advantages. For example, SKA sky surveys will include billions of sources, and each source will be described and characterized by many different parameters including location, size, morphology, flux density, spectral index, line width, redshift etc. Combining this information with comparable surveys at many other wavelengths (a veritable astronomy mashup), requires a new approach to large-scale data analysis, and there can be little doubt that the techniques now being pursued commercially in the realm of Big Data can also be applicable to the astronomy case. In addition to advanced pattern recognition, algorithms working on streaming data can be specifically designed to identify anomalies or outliers, and to preserve them (rather than the case now where they are often thrown away). This can be important in enabling serendipitous discovery relevant to a wide-range of studies but perhaps Transients and SETI (Search for Extraterrestrial Intelligence) research stand to benefit most.

Another area where commercial investment in data analytics is making great strides is in data visualization. Progress in data visualisation has been at a virtual standstill for the last few decades in astronomy, for example, spectral line cube data sets remain cumbersome to inspect and explore. With data analytics we can hope to use new visualization approaches, featuring automatic feature extraction and collaborative sharing tools. Finally, predictive analytics (and in principle also prescriptive analytics) must also have something to teach astronomers, especially in the area of regular variables or repeaters, and again transients.

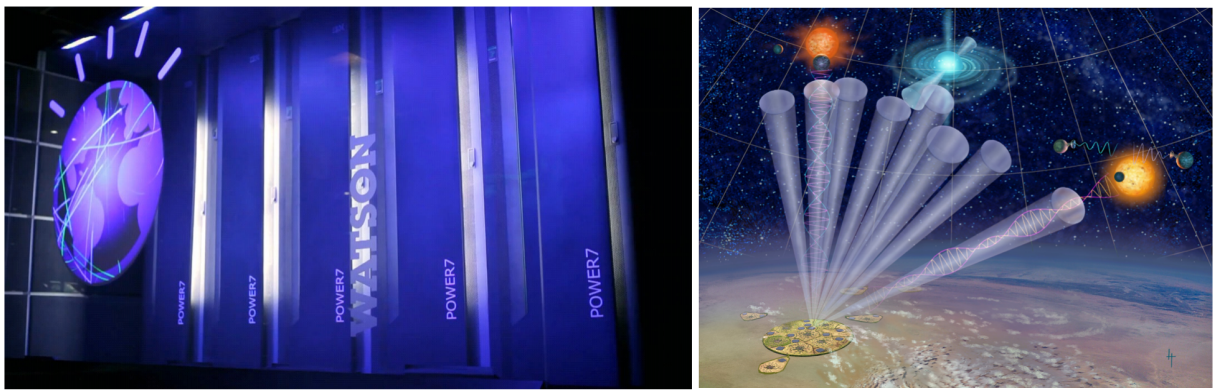

**Figure 1: Applying cognitive computing to large astronomical surveys could have a huge impact in terms of serendipitous discovery - areas that might specifically benefit include those like SETI where human bias and other pre-conceptions may limit current efforts.**

## 3. Cognitive computing for astronomy

Cognitive computing seeks to leap beyond the deterministic boundaries of conventional programmable (von Neumann) computers by introducing dynamic systems that learn from the data, sense their environment, place unstructured data such as video in context, independently pursue new courses of action and directly interact with humans by understanding and using natural language. IBM's Watson machine [6] represents an important step forwards towards this goal but further progress requires revolutionary new architectures to be developed that mimic the human brain in which memory and processing power are tightly integrated into sub-units that are distributed, run in parallel and communicate (both internally and externally) via nano-photonic connectors [7].

The impact of such systems, guided by experienced scientists would surely be profound. At least initially, the astronomer's human qualities of intuition, insight and creativity would be important guiding inputs to the direction this new discovery process might take. Many fields of science stand to benefit, especially those that are data dominated. Naturally there are also risks involved in developing technologies that approach or perhaps even exceed human capacity. But this is the slippery path on which we are all travelling – natural and good communication between machines and humans will be vital element towards a successful outcome.

## 4. Conclusions

Over the last decade, radio astronomy has successfully embraced COTS (Commercial off-the-shelf) technology with the new generation of telescopes like LOFAR making good use of the many advances fuelled by the global investment in ICT. The huge volume and increased complexity of future large sky surveys, coupled with the emergence of multi-messenger facilities, demand a new approach to scientific discovery in astronomy (but also other fields). An opportunity exists to make great leaps forward by engaging with the Big Data programme, re-using the rapidly developing suite of data analytics (and the associated visualisation techniques). This approach can greatly benefit the full scientific exploitation of large astronomical surveys, leading to new discoveries on accelerated timescales.

The great promise of cognitive computing is also on the horizon – this could propel astronomical research in very new and perhaps unexpected directions. Projects like DOME are important vehicles for ensuring early engagement of the astronomy community with cognitive systems. In particular, areas such as SETI research might really benefit from an approach that is largely independent of human preconceptions, potentially leading to game-changing discoveries for all mankind.

**Acknowledgments**

I would like to acknowledge the DOME project. Attendance of the author at Radio 2014 was partially supported via an IBM Faculty Award.